\PassOptionsToPackage{svgnames}{xcolor}
\documentclass[journal,hideappendix]{vgtc} 

\onlineid{1789}

\vgtccategory{Research}

\vgtcpapertype{Theoretical \& Empirical }

\title{The Impact of Visual Segmentation on Lexical Word Recognition}

\author{%
  \authororcid{Matthew Termuende}{0009-0000-9269-7703},
  Kevin Larson,
  \authororcid{Miguel Nacenta}{0000-0002-9864-9654} 
}

\authorfooter{
  \item
  	Matthew Termuende is with the University of Victoria.
  	E-mail: mtermuende@uvic.ca.
  \item
  	Kevin Larson is with Advanced Reading Technologies at Microsoft Research.
  	E-mail: kevlar@microsoft.com.

  \item Miguel Nacenta is with the University of Victoria.
  	E-mail: nacenta@uvic.ca.
}


\teaser{
  \centering
  \includegraphics[width=\linewidth, alt={Figure shows two example sentences showcasing two of the interventions tested in the paper. The sentence "A pharmacological catastrophe caused upheaval" is written above with phoneme units grouped by alternating the text color between black and red. The sentence "The unionized workers refused to use the unionized compound" is written below with outlines around syllables.}]{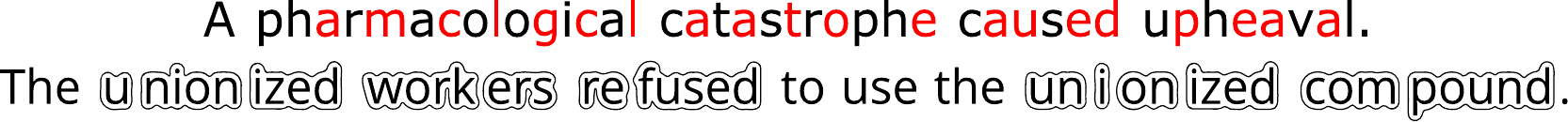}
  \caption{
  Example sentences with interventions that visually segment phonemes (top) and syllables (bottom).
  }
  \label{fig:teaser}
}

\graphicspath{{figs/}{figures/}{pictures/}{images/}{./}} 

\usepackage{mathptmx}                  

\usepackage{todonotes}
\usepackage{enumitem} 
\usepackage[font=normalsize]{subcaption} 
\usepackage{colortbl}
\usepackage{xcolor}
\usepackage{amssymb}
\usepackage{xspace}
\usepackage{float}
\usepackage{booktabs}
 \usepackage{amsmath}
 \usepackage[normalem]{ulem}
 \usepackage{tcolorbox} 
 \usepackage{caption}

\definecolor{strongSupport}{RGB}{109, 184, 116}
\definecolor{weakSupport}{RGB}{173, 193, 79}
\definecolor{strongReject}{RGB}{232, 136, 86}
\definecolor{weakReject}{RGB}{229, 177, 69}

\DeclareUnicodeCharacter{FF1F}{\textquestiondown} 

\newcommand{\inlinesection}[1]{\vspace{0.25em}\noindent$\blacktriangleright$~\textsc{\textbf{#1}}\xspace}

\abstract{
When a reader encounters a word in English, they split the word into smaller orthographic units in the process of recognizing its meaning.
For example, ``rough'', when split according to phonemes, is decomposed as r-ou-gh (not as r-o-ugh or r-ough), where each group of letters corresponds to a sound. Since there are many ways to segment a group of letters, this constitutes a computational operation that has to be solved by the reading brain, many times per minute, in order to achieve the recognition of words in text necessary for reading. In English, the irregular relationships between groups of letters and sounds, and the wide variety of possible groupings make this operation harder than in more regular languages such as Italian.
If this segmentation takes a significant amount of time in the process of recognizing a word, it is conceivable that providing segmentation information in the text itself could help the reading process by reducing its computational cost. In this paper we explore whether and how different visual interventions from the visualization literature could communicate segmentation information for reading and word recognition. We ran a series of pre-registered lexical decision experiments with 192 participants that tested five main types of visual segmentations: outlines, spacing, connections, underlines and color. The evidence indicates that, even with a moderate amount of training, these visual interventions always slow down word identification, but each to a different extent (between 32.7ms---color technique---and 70.7ms---connection technique). These findings are important because they indicate that, at least for typical adult readers with a moderate amount of specific training in these visual interventions, accelerating the lexical decision task is unlikely. Importantly, the results also offer an empirical measurement of the cost of a common set of visual manipulations of text, which can be useful for practitioners seeking to visualize alongside or within text without impacting reading performance. Finally, the interaction between typographically encoded information and visual variables presented unique patterns that deviate from existing theories, suggesting new directions for future inquiry.
}
\begin{document}
\let\ab\allowbreak 


\newif\ifdiffversion
\diffversionfalse 

\newtcolorbox{diffnewfigurebox}{colback=White, colframe=ForestGreen, 
  boxrule=1pt, arc=3pt, left=4pt, right=4pt, top=4pt, bottom=4pt}
\newtcolorbox{diffoldfigurebox}{colback=White, colframe=Maroon, 
  boxrule=1pt, arc=3pt, left=4pt, right=4pt, top=4pt, bottom=4pt}

\newcommand{\diffold}[1]{%
  \ifdiffversion
    \textcolor{Maroon}{\sout{#1}}%
  \else
  \fi
}

\newcommand{\diffnew}[1]{%
  \ifdiffversion
    \textcolor{ForestGreen}{#1}%
  \else
    #1
  \fi
}

\newcommand{\removed}[1]{%
  \ifdiffversion
    \diffold{#1}%
  \else%
  \fi%
}%

\newcommand{\diffoldfig}[1]{%
  \ifdiffversion
    \begin{diffoldfigurebox}
    #1
    \end{diffoldfigurebox}
  \else
  \fi
}

\newcommand{\diffnewfig}[1]{%
  \ifdiffversion
    \begin{diffnewfigurebox}
    #1
    \end{diffnewfigurebox}
  \else
    #1
  \fi
}

\maketitle
\section{Introduction}
According to Saenger\cite{saenger1997space}, the addition of space between words in written text was a pivotal graphical innovation that led to faster, silent reading. 
This disrupted the established tradition of "scriptura continua," in which sentences were written without spaces. Nowadays English and many other languages clearly delineate words with spaces.
Empirical evidence supports that unspaced text is read more slowly, and word identification within such text is more error-prone~\cite{mirault_reading_2019}. 

\begin{figure}[h]
    \centering
    \includegraphics[width=0.5\linewidth, alt={Figure shows the heirarchy of phonological unit granularities: syllables, made of onset-rimes, made of phonemes. The word "photograph" is printed, and its transription in international phonetic alphabet characters is printed above the word. Colored boxes are drawn within the transcribed word illustrating grain sizes. A box labeled "syllable" encloses "graph"; boxes labeled "onset" and "rime" enclose "g r" and "\ae f" respectively; a box labeled "phoneme" encloses "f".}]{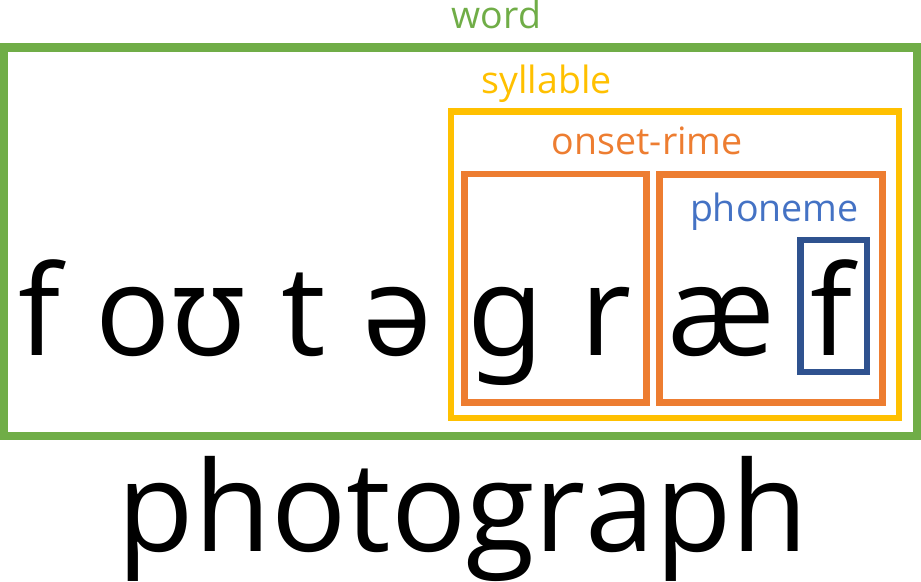}
    \caption{The word "photograph" (in IPA~\cite{international1999handbook}) shows the hierarchy of unit granularities: syllables, made of onset-rimes, made of phonemes.}
    \label{fig:phonGrainSizes}
\end{figure}

Beyond words, reading involves smaller units (e.g., syllables and phonemes) which are not visually salient in normal text~\cite{goswami2003nonword}.
Spoken words are structured hierarchically: words are composed of syllables, which divide into onset and rime, and further into phonemes—--the smallest meaning-distinguishing phonological units (see Figure~\ref{fig:phonGrainSizes}). In alphabetic writing systems like English, these phonological units correspond to orthographic (written) units. 
Reading acquisition involves learning to map between these phonological and orthographic units and
ability to blend and segment phonemes early in life predicts reading ability~\cite{ball1991does,clayton2020longitudinal}, 
highlighting the importance of sub-word units in literacy.

Given the profound effect of the visual intervention of word spacing in reading, and the existing evidence of the importance of sub-word segmentation, it is plausible that visual interventions at smaller granularity (e.g., syllable or phoneme) could further support reading. Although research has shown successful related visual interventions of other types (e.g., typographically highlighting grammatical structure speeds up and increases comprehension~\cite{north1951reading} and altering typeface according to intonation improves vocal expression of children reading aloud~\cite{bessemans_visual_2019}), we do not have conclusive evidence of whether or how making sub-word segmentation visible can improve word recognition.


To address this lack of evidence we carried out a series of experiments that measure lexical decision time with a broad set of visual ways to make segments visible: color, connection, outline, underline and spacing.
This is the first empirical data comparing these interventions, as well as two different levels of granularity (syllables and phonemes). We hypothesized that such enhancements would lead to faster and more accurate word identification. Contrary to our expectations, the results show that changing the visual appearance of text, even if it is to provide potentially useful information, slows down the reading process. A second and third experiment were designed to rule out effects of the experimental design and to test whether instruction and training reverse or mitigate the results. We found that even with instruction and training, the visual interventions still impede word recognition. 

The results are useful in two ways. First, they show that speeding up reading with segmentation information provided visually is harder than expected. Second, they provide a ranking and measures of the extent to which different visual interventions affect word recognition, which can be useful for visualization practitioners who encode data in text.

\section{Background and Related Work}
We first provide some conceptual background before situating our research with respect to important related work.

\subsection{Word Identification and Lexical Decision}
\label{sec:related:wordIdentification}
Word identification (or word recognition)---associating the written word with its meaning---is central to reading~\cite{rayner2012psychology}. It is largely the same in isolation as in connected text, with each word identified upon fixation, although there are effects of context~\cite{rayner2012psychology}. Depending on the paradigm, experiments of isolated word identification assume the reader has recognized a word if they can name it (naming), make a categorical judgment based on its meaning (categorization), or decide whether it is a word or a \emph{pseudoword}---a meaningless string of letters that conforms to the patterns of the language (lexical decision---LD). Measures of performance in these tasks include response latency and accuracy. These tasks present varying decision and motor demands\cite{carreiras2007Brain,rayner2012psychology}, and exhibit different sensitivity to linguistic variables (such as frequency of occurrence)~\cite{balota2004visual, balota1984lexical}, but show similar patterns of brain activation in areas associated with lexical retrieval \cite{carreiras2007Brain}. 

\subsection{Phonology and Orthographic Units}
\label{sec:backgroundOrhographicUnits}
English words have a hierarchical structure of phonological units (units of sound): words are composed of syllables, which divide into onset-rimes, and further into phonemes, the smallest meaning-distinguishing unit of sound~\cite{treiman2017role} (see Figure~\ref{fig:phonGrainSizes}). In written English, these correspond systematically to sequences of letters (orthographic units), with larger grain sized units tending to be more predictive of correct pronunciation~\cite{ziegler_reading_2005}. English is a \emph{deep} or \emph{inconsistent} orthography (written language), meaning that the mappings between letters and phonemes are highly dependent on context~\cite{borleffs_measuring_2017}.

Multiple lines of evidence underscore the importance of orthographic units in reading. Word identification performance is affected by priming with syllables~\cite{ferrand1997syllable} or rimes~\cite{Bowey1990Orthographic}, and exhibits effects of frequency of component syllables~\cite{hutzler2005effects,macizo2007syllable} and onset-rimes~\cite{bowey1994development}. Cross-language comparisons suggest that orthographic consistency influences preferred orthographic unit granularity in word identification~\cite{ziegler2001identical,goswami2003nonword}, and is associated with differences in the manifestation of dyslexia~\cite{ziegler_reading_2005}. Inconsistent orthographies are associated with slower reading acquisition~\cite{ziegler_reading_2005}.

\subsection{Relating Visual Segmentation to Text Visualization}
\label{sec:textVis}
Typographic parameters such as letter shape or spatial arrangement can be used as \emph{typographic visual variables} to represent or \emph{encode} data using text as a \emph{visual mark}~\cite{brath2019bertin,brath2020visualizing,lang2022perception}. If our purpose were to visually encode the nominal phonological data (the specific phonemes or syllables) using the orthographic units as visual marks, our work could be guided by existing text visualization frameworks (e.g., see \cite{brath2020visualizing}). However, the objective of our \emph{visual segmentation} interventions is to visually communicate the decomposition of words into orthographic units, which instead corresponds to demarcating the physical marks (group of glyphs) associated with each data case. To our knowledge, this is an underexplored problem in text visualization.

\subsection{Enhancing Reading using Typographic Variables}
\label{sec:relatedWork:enhancingTypographicVariables}
Encoding linguistic information onto text enhances reading performance for a variety of demographics, from children~\cite{bessemans_visual_2019} to skilled readers~\cite{north1951reading,graf1966perception,gu_ai-resilient_2024}. Bessemans et al.~\cite{bessemans_visual_2019} used boldness, letter width and vertical position to communicate volume, duration and pitch within words. They found that, only when instructed on the encodings, children were able to use them to read aloud with greater vocal inflection.
A case study of one adult with severe dyslexia found improved speed and accuracy on a naming task using hyphens to mark syllables~\cite{harley_hyphenation_2006}.
Visual Syntactic Formatting uses spaces or line returns to organize words into phrases within sentences. It has been empirically proven to improve reading speeds by over 10 percent and to enhance the accuracy of answers to comprehension questions by over 15 percent~\cite{north1951reading,graf1966perception}.
Gu et al.~\cite{gu_ai-resilient_2024} presented a technique to typographically signal the importance of words within sentences, finding benefits to comprehension.

\subsection{Pedagogical Interventions Using Visual Segmentation}

Typographic interventions that segment words into orthographic units have been investigated to facilitate reading acquisition. 
Training with a \emph{cued orthography} that segments words into graphemes is found to help early readers to correctly construct words from grapheme segments~\cite{oliver_differentiation_1972,oliver_transfer_1973}. Some authors have developed cued orthography interventions that also disambiguate the grapheme-phoneme mapping using recognizable icons~\cite{donnelly_annotating_2020,seward_linking_2014,seward_impacting_2024}. Donelly et al.~\cite{donnelly_annotating_2020} found that struggling readers improved in rate and accuracy of verbal reading after take-home practice with such an intervention.

Visual segmentation is similar to a pedagogical technique called \emph{Visual Input Enhancement - VIE}~\cite{smith_input_1993}. VIE uses \emph{typographic cues} to emphasize linguistic units, helping second language learners notice the patterns of the language~\cite{lee2006synthesis}. Findings regarding efficacy conflict with each other~\cite{han_textual_2008}, but there is consensus on its positive effect on learning and negative effect on comprehension (for reviews see~\cite{lee2006synthesis,lee_visual_2008,han_textual_2008}).

Evaluations of related pedagogical interventions such as VIE focus on reading acquisition, and the demographics (learners) and measures used (e.g., measures of learning, such as sentence completion~\cite{han_textual_2008}) are not informative for skilled word identification. 

\subsection{Psycholinguistic Studies of Orthographic Processing}
\label{sec:relatedSegmentationStudies}
The empirical work closest to our own are psycholinguistic studies that test predictions of psycholinguistic theory. These studies have used experimental paradigms 
such as lexical decision (LD)~\cite{katz_linguistic_1981,katz_syllable_1983,nilsson_insertion_1990,levitt1991syllable,treiman1994extent,pring1981phonological}, go-no-go lexical decision~\cite{katz_syllable_1983}\diffold{(GNG-LD)}, verbal lexical decision~\cite{katz_linguistic_1981}\diffold{(V-LD)}, naming~\cite{bowey1996orthographic,katz_linguistic_1981}, passage reading~\cite{nilsson_insertion_1990}\diffold{(PR)}, and pseudoword forced-choice~\cite{treiman1994extent}\diffold{(PW-FC)}. Interventions have used a variety of typographic cues, including forward slashes~\cite{katz_linguistic_1981,katz_syllable_1983, treiman1994extent}, asterisks~\cite{nilsson_insertion_1990,levitt1991syllable}, spaces~\cite{nilsson_insertion_1990}, and case or script~\cite{katz_linguistic_1981,bowey1996orthographic}. Two studies investigated multiple encodings~\cite{katz_linguistic_1981,nilsson_insertion_1990}, but none have explored more than two alternatives. Orthographic units have been investigated in several grain sizes, including syllables~\cite{katz_linguistic_1981,katz_syllable_1983,nilsson_insertion_1990}, onset-rimes~\cite{levitt1991syllable,treiman1994extent,bowey1996orthographic}, and phonemes~\cite{pring1981phonological}. To our knowledge, multiple grain sizes have not been compared in a single study. 

Since experimental paradigm, typographic design, and orthographic grain size vary between studies but seldom within studies, it is difficult to reconcile the disparate findings. We review them here, organized by orthographic grain size.

\noindent \textbf{Syllable Segmentations.}
Children perform faster and more accurately in naming and\diffold{V-LD} \diffnew{verbal LD}~\cite{katz_linguistic_1981} and less-skilled children perform faster in LD~\cite{katz_syllable_1983} when syllables were correctly vs.\ incorrectly segmented using forward slashes (e.g., ex/am/ple). With the same encodings, adults perform faster in\diffold{GNG-LD} \diffnew{go-no-go LD} for correct vs.\ incorrect segmentations when the task was made more difficult by alternating case (e.g., eX/aM/pLe)~\cite{katz_syllable_1983}; results were more equivocal without alternating case. Adults were faster in \diffold{PR} \diffnew{passage reading} with syllables correctly vs.\ incorrectly segmented using spaces (e.g., ex am ple), but the opposite was found using asterisks (e.g., ex*am*ple)~\cite{nilsson_insertion_1990}. For one of these encodings (it is unclear from the text whether spaces or asterisks were used), there is evidence that correct outperform incorrect segmentations in LD, and that both are worse than plaintext~\cite{nilsson_insertion_1990}. The only study evaluating training effects (in LD) concluded that results are unlikely to change with training~\cite{nilsson_insertion_1990}; however, methodological problems advise caution. Adults with a more shallow orthography (Serbo-Croatian) exhibit faster LD responses for correct vs.\ incorrect syllable segmentations when syllables are indicated by forward slashes, but not when indicated by script (similar to case in English)~\cite{katz_linguistic_1981}. 

\noindent \textbf{Onset-rime Segmentations.}
With asterisk as the segmentation cue, correct onset-rime segmentations elicited fewer errors in LD than rime-disturbing incorrect segmentation, but not onset-disturbing segmentation~\cite{levitt1991syllable}.
With double forward slashes as cue (e.g., gr//aph), faster LD and \diffold{PW-FC} \diffnew{pseudoword forced-choice} responses were observed for correct vs.\ rime-disturbing segmentations~\cite{treiman1994extent}. With case as cue, word naming performance was faster for correct vs.\ onset-interrupting segmentations, but not rime-interrupting segmentations~\cite{bowey1996orthographic}; however, the effect disappeared when case was reversed (e.g., shRED instead of SHred).

\noindent \textbf{Phoneme Segmentations.}
One study involving phoneme segmentations found no difference on LD between correct vs.\ incorrect segmentations using case~\cite{pring1981phonological}. However, their finding that the correct and incorrect segmentations respectively preserved or disturbed the pseudohomonym effect (i.e., it takes longer to reject pseudowords that are pronounced as an existing word, e.g., phast) suggests that phoneme segmentations interact with phonological processing during reading. 

\subsection{Research Gaps and Significance}
The prior psycholinguistic studies discussed above do not conclusively assess the potential for enhancement of word identification, as their goal is instead to test predictions of theory. A rigorous comparison with a plaintext condition is missing, as is an evaluation of effects of training and instruction. Effects of segmentation information are unclear: it sometimes enhances (relative to meaningless controls) or has no effect; this seems to vary with graphical designs and phonological grain sizes, but may be due to differences in study design. While there is some indication that choice of graphical design impacts word identification, there is no comparison of performance between different graphical designs to inform visualization design. Our work addresses these gaps.
\section{Research Goals}
\label{sec:goals}
Our work has a primary and a secondary research goal. The primary goal is to assess whether graphical segmentation interventions at the sub-word level (e.g., spacing between syllables) have the potential to enhance word recognition for skilled readers. We use lexical word recognition time (see Section~\ref{sec:exp1:task} for a full description of the task) as a proxy for reading speed. Although lexical recognition might not reflect all aspects of reading (see Section~\ref{sec:related:wordIdentification}), showing speed-ups would be a strong indicator of the possibility of speeding up reading itself. 

The secondary goal is to compare different graphical interventions in terms of their effect on the lexical recognition task (also as a proxy for effects on reading). An assessment of which interventions are more intrusive on the reading process can be useful for designers who want to overload text renderings with additional information, such as in visualization (e.g.,~\cite{brath2020visualizing,lang2022perception}) or for graphical or aesthetic purposes.

\section{Intervention Design} \label{sec:intervention_design}

The objective of visual segmentation interventions is to visually communicate the decomposition of words into orthographic units. For syllables and phonemes this means grouping several letters that correspond to a syllable (or phoneme) in a visual unit that makes it distinguishable from the other units. Inspired by Gestalt principles of perceptual grouping~\cite{wagemans2012century}, we suggest that letters may be grouped into segments through similarity, proximity, connectedness, or common region.

\begin{figure}
  \centering
  \subfloat[No Intervention]{\label{fig:exampleNone}\includegraphics[width=0.33\linewidth]{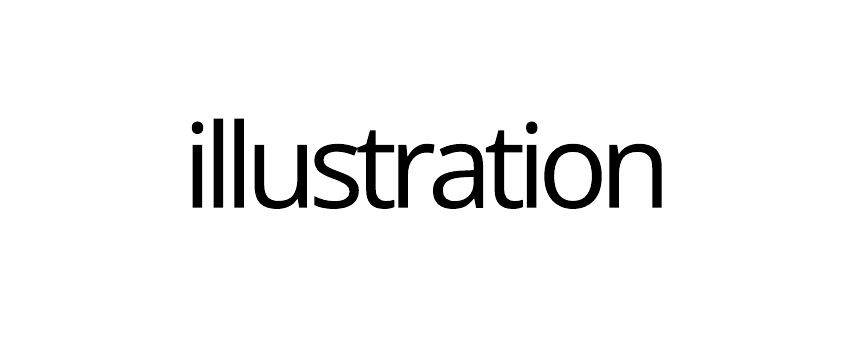}}
  \hfill
  \subfloat[Color]{\label{fig:exampleColor}\includegraphics[width=0.33\linewidth]{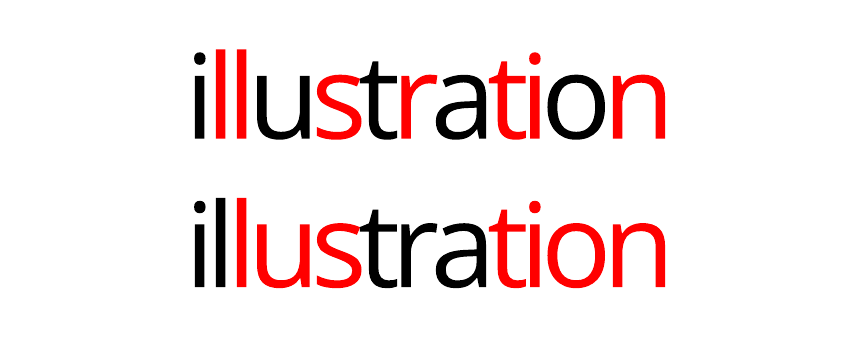}}
  \hfill
  \subfloat[Connection]{\label{fig:exampleConnections}\includegraphics[width=0.33\linewidth]{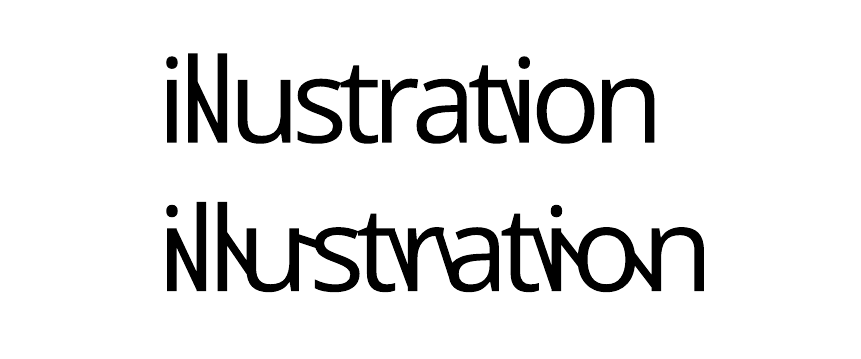}}
  
  \subfloat[Outline]{\label{fig:exampleOutlines}\includegraphics[width=0.33\linewidth]{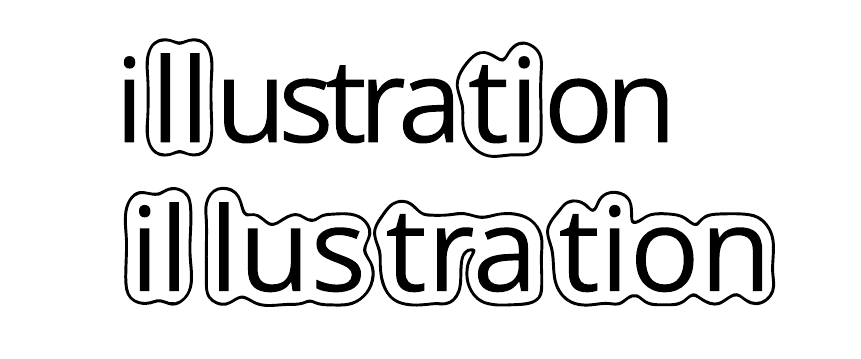}}
  \hfill
  \subfloat[Underline]{\label{fig:exampleUnderlines}\includegraphics[width=0.33\linewidth]{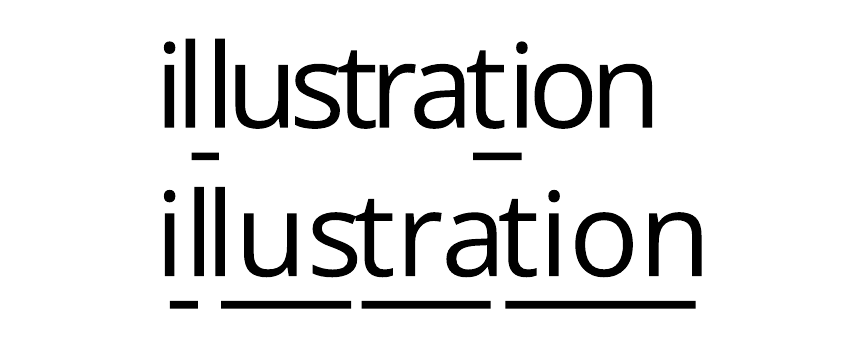}}
  \hfill
  \subfloat[Spacing]{\label{fig:exampleSpacing}\includegraphics[width=0.33\linewidth]{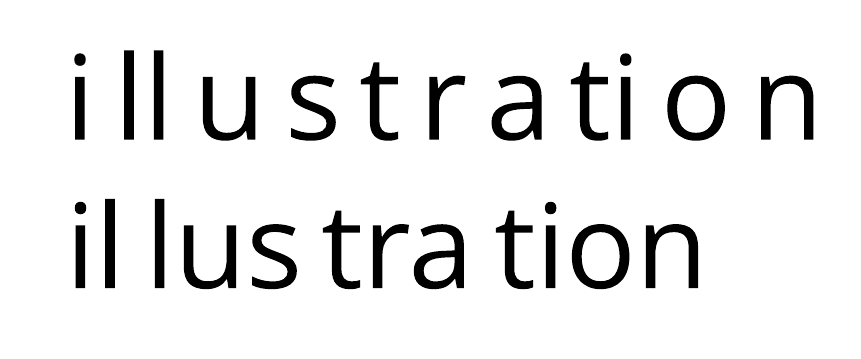}}
  \caption{The selected visual segmentation designs for phoneme (top) and syllable (bottom) grain sizes.} 
  \label{fig:designExamples}
\end{figure}

We reasoned that our research goals would be best supported by taking a broad sample of the design space. This is underscored by the heterogeneous findings of prior studies (see Section~\ref{sec:relatedSegmentationStudies}). Considering practical constraints on budget and experiment duration, we selected five graphical designs that represent a variety of conventional \diffold{uses and expected interactions with reading} \diffnew{and non-conventional ways to group letters}. We selected phonemes and syllables as target orthographic units because they are the finest and largest grained phonologically meaningful sublexical segments respectively. Figure~\ref{fig:designExamples} shows our choices of graphical designs.

\label{sec:visualDesign}

\inlinesection{Color} (Figure~\ref{fig:exampleColor}). We alternated the rgb color of segments between black ($rgb(0,0,0)$) and red ($rgb(1,0,0)$). Since color hue and luminance detection are separated early in the visual system~\cite{hering1964outlines}, segmentation information presented as a color signal might interact minimally with letter recognition that uses a luminance signal. Color is an example of a design that uses similarity: grouping letters that share a common visual or typographic attribute (such as those previously studied for perceptual qualities such as discriminability~\cite{lang2022perception,Strobelt2016guidelines,brath2020visualizing}).

\inlinesection{Connection} (Figure~\ref{fig:exampleConnections}). We connected letters within segments by adding straight line strokes with a width of 60\% of the font stem width (FSW--- the width of an 'l' glyph). We manually selected junction locations that result in downward-angled strokes, which are rare in the Latin script. We reasoned that this would make them more noticeable. As ligatures connecting letters for aesthetic reasons (such as fi) are commonplace and have little effect on processing of letters in word identification~\cite{fernandez2023breaking}, it is conceivable that connection may be a minimally disruptive intervention. However, there are likely perceptually better or worse design decisions, as some geometry and configurations of strokes seem to be preferred in natural writing systems~\cite{changizi2006structures}.

\inlinesection{Outline} (Figure~\ref{fig:exampleOutlines}). We created the outlines by applying a heat diffusion algorithm on the original segment glyphs and fitting a spline to the diffused boundary, drawn with a thickness of 40\% FSW. This surrounds segments with a boundary that conforms to the rough shape of the letters within the segment. We reasoned that this is the most natural (most similar to the original glyphs) example of a design that groups letters by common region. 

\inlinesection{Underline} (Figure~\ref{fig:exampleUnderlines}). A single underline was drawn under each segment, 3 FSW below the baseline, with a width calculated to leave at least 1 FSW between subsequent underlines and a thickness of 80\% FSW. This is another design that groups by common region, but one in which only part of the boundary is drawn. As the new marks introduced are outside the bounding box of the original text, it is likely less obtrusive than other annotation alternatives. On the other hand, this same unobtrusiveness may make it less likely to effectively communicate the information. This is also a common design used in pedagogical interventions, lending it ecological relevance.

\inlinesection{Spacing} (Figure~\ref{fig:exampleSpacing}). The perceptual tendency to group proximal objects can be used to segment words by reducing spaces between letters within segments or by increasing the distance between segments. However, inter-letter distance is strongly associated with the phenomenon of visual crowding~\cite{pelli2008uncrowded}, which disrupts reading~\cite{pelli2007crowding}, especially for people with dyslexia~\cite{martelli2009crowding}. We therefore chose to increase space between segments for this design. We matched the distances between segments to those in the outline design. Graphically, spacing is minimally invasive as it does not introduce or alter marks. However spaces are also used to mark letter and word boundaries, and it is unclear how this might interact with segmentation into orthographic units.
\section{Research and Analysis Approach}
We use a controlled experimental approach with remote participation of human participants over a web platform~\cite{heerCrowdsourcingGraphicalPerception2010}. 
Remote participation offers less control over the experimental environment because participants carry out the experiment in their own computer at their location of choice; however, we ameliorate as much of of the variability as possible (e.g., by controlling the size of the stimuli, see~\ref{sec:exp1:task}). Natural variation in the environments and participant circumstances is also not all bad: it provides some ecological validity since variation in the participant environments will likely resemble the varied and naturalistic reading environments and circumstances.

We initially designed one experiment. Based on the results, we iterated to rule out alternative explanations for the data. All experiments were pre-registered (precludes ``fishing for results'' and HARKing~\cite{cockburnHARKNoMore2018}). The terms in the pre-registrations vary with respect to those in the paper for presentation and clarity purposes. Non pre-registered statistical tests are marked as exploratory. All deviations with respect to the planned pre-registered analysis are stated where appropriate.   

We apply a Markov-Chain Monte Carlo (MCMC) Bayesian approach to data analysis~\cite{kruschkeDoingBayesianData2014,gelmanBayesianDataAnalysis2013}. The Bayesian approach allows us more flexibility when building the models, supports easier interpretation of the results and avoids some of the pitfalls of interpreting p-values~\cite{cummingNewStatisticsWhy2014,klineSignificanceTestingStatistics2013}. We carefully selected uninformative priors and checked all chains for good mixture, making sure that equivalent sample sizes are $> 10,000$. The MCMC simulations were implemented using JAGS version 4.3.0~\cite{jags}, using the R-JAGS interface~\cite{rjags} running on R version 4.1.3~\cite{R}.

We use a parameter-estimation approach to hypothesis testing\cite[Chapter~12]{kruschkeDoingBayesianData2014}. Our analyses estimate the credible values of model parameters for which we report the $95\%$ credible interval and the mean estimate. Our hypotheses involve statements about these parameters (e.g., that the difference between two specific parameters is greater than zero). When we report probability values, this is the probability, given the data, that the hypothesis is true. \diffnew{The probabilities are calculated by subtracting the posterior samples of the conditions under comparison and reporting the proportion of samples where one condition is larger than the other. As the posterior samples approximate the posterior distribution given the data~\cite[Chapter~7]{kruschkeDoingBayesianData2014}, we interpret the proportion of samples satisfying a hypothesis as approximating the probability mass of the posterior distribution for which the hypothesis is supported.} There is some similarity to significance testing, but the reported values are not identical to frequentist p-values~\cite[Chapter~11]{kruschkeDoingBayesianData2014}. Measures associated with a model-comparison approach (e.g., Bayes Factor) are not relevant to our approach.

\diffnew{
We use Bayesian models with a metric predicted variable (log-transformed response time). Each hypothesis involves a specific slice of the dataset. We model all factors that vary in the data slice in question as nominal predictors. Depending on the hypothesis, this may only be a single factor (e.g., Experiment 1, H2.2 models meaningfulness), or it may be all possible factors (e.g. Experiment 1, H0.2 models stimulus type, segmentation type, meaningfulness and visual design). We also model all possible interactions between factors, and include a non-interacting random (participant) factor in each model. We model the predicted variable as Student-t distributed, with mean determined by the sum of effects of the nominal factors, which we model as normally distributed with gamma-distributed priors for variance. The exact models and analysis scripts are included in the supplemental materials.}

\diffold{Where applicable, we report Cohen's d value as a measure of effect size. We calculate Cohen's d associated with a difference between parameter values, for each sample of the MCMC simulation, using the pooled standard deviation for all parameters, and report the mean value.} 
\section{Experiment 1: Impact of Visual Segmentation on Lexical Decision}
To address the goals described in Section~\ref{sec:goals} we designed an experiment using the forced-choice lexical decision paradigm.\footnote{The pre-registration of this study is available at \url{https://doi.org/10.17605/OSF.IO/RG7FH}.}

\subsection{Task and Measurement}
\label{sec:exp1:task}
In the lexical decision (LD) task the participant must decide whether the text shown on screen is a word (e.g., perception) or a pseudoword (e.g., percestion). In each trial, participants looked at a fixation mark (crosshairs). When the stimulus (word or pseudoword) replaced the crosshair, the participant pressed the ``A'' key if they recognized the stimulus as a word or ``L'' if not. The main measure is elapsed time between stimulus onset and key press. The stimulus was rendered in the Open Sans font with a fixed size (vertical subtended angle of 0.646 degrees, approximately 16pt font at 50cm). The LD reaction time is widely used as a proxy for word identification speed (see~\cite[p.51]{rayner2012psychology} for a discussion). Time is measured in milliseconds and log-transformed for analysis. Reported results (e.g., means and high-probability density intervals) are back-transformed from the logarithmic domain.

\subsection{Experimental Factors \diffold{and Conditions}}
The experiment manipulated four factors. The first one is \emph{stimulus type} (each trial could be a word or a pseudoword). Trials with a pseudoword make the task non-trivial. The other three factors together determine the intervention. In the following description of the factors we do not describe the baseline, which is just showing text unaltered (hereafter we refer to this as the \emph{no-intervention} or \emph{none} condition).

\inlinesection{Segmentation Type:}
    The phonological unit used to break the word up into segments (\textbf{phonemes} or \textbf{syllables}). This factor tests the effect of phonological grain size (see Section~\ref{sec:backgroundOrhographicUnits}).
    
\inlinesection{Visual Segmentation Design:}
    The way in which the different segments are visually differentiated from each other. \textbf{(spacing, underline, color, outline, connection, none)}.
    Displayed in Figure~\ref{fig:designExamples}, this factor tests the range of possibilities described in Section~\ref{sec:visualDesign} and also allows us to address the secondary research goal (see Section~\ref{sec:goals}). 
    
\inlinesection{Meaningfulness:}
    Whether the segmentation is correct with respect to the unit of segmentation or not (\textbf{meaningful} or \textbf{meaningless)}. Meaningless segmentations use incorrect separation of units (e.g., e-xa-mple). This helps distinguish between the effects of real segmentation information and effects of the visual intervention. 

\subsection{Research Questions and Hypotheses}\label{sec:exp1:RQs}
To address our primary research goal 
of assessing the enhancement potential of segmentation interventions, we focused on the following research questions (RQs):\footnote{Reference names for RQs in the paper and the pre-registration differ due to clarity and exposition reasons.}
\begin{description}
\item[\texttt{RQ1}:] Can word identification be enhanced by visually encoding phonologically meaningful segmentation information?
\item[\texttt{RQ2}:] Is the information communicated by visual segmentation responsible for any effect on word identification?
\item[\texttt{RQ3}:] Does the phonological granularity of visually presented segmentation information affect word recognition performance?
\end{description}
In connection with our secondary research goal
of comparing the effects of the graphical forms on reading, we investigated the following RQs:
\begin{description}
\item[\texttt{RQ4}:] Can word identification be impeded by phonologically meaningless visual segmentation?
\item[\texttt{RQ5}:] Which interventions have a greater or lesser impact on word identification?
\end{description}

Each research question is addressed by one or more hypotheses, which have their corresponding statistical tests. For clarity and brevity, we do not report results for each tested hypothesis here. Instead, we provide the formal definitions of hypotheses and their corresponding test results for this and the subsequent experiments in \iflabelexists{appendix:hypotheses}
  {\cref{appendix:hypotheses}}
  {the Appendices file of the supplemental materials}.

\subsection{Experimental Design and Conditions}
We used a fully-crossed within-subjects factorial design. 
Participants completed a total of 500 trials: 20 training repetitions and 10 repetitions in each of the 48 different factor combinations: 2 (word/pseudoword) $\times$ 2 (\emph{segmentation type}) $\times$ 6 (design and no-intervention) $\times$ 2 (meaningfulness) $= 48$ combinations. The training used no visual intervention (10 word, 10 pseudoword) and was excluded from  analysis. Note that meaningfulness and segmentation type are irrelevant for the baseline condition (text without segmentation does not have a meaningful/meaningless or syllable/phoneme segmentation). Thus, the baseline condition was performed four times as often as all other combinations. 

We blocked stimuli by \emph{visual design}, counterbalanced by a digram-balanced Latin square. Within each \emph{visual design} block, we further blocked by \emph{segmentation meaningfulness}, with alternating levels (either meaningful or meaningless first in every block). Within \emph{segmentation meaningfulness} blocks, we further blocked by \emph{segmentation type} with alternating levels. This results in 6 $\times$ 2 $\times$ 2 $= 24$ different orderings that balance condition order in all combinations of the factors and sequential effects between \emph{visual design} blocks. We intended to assign two participants to each of the 24 possible orderings for complete counterbalancing. However, due to a technical problem with condition assignment, four of the orderings were each assigned three participants, and another four orderings were only each assigned one participant.

\subsection{Participants}
We recruited adult participants ($>18 yo$) online using the Prolific platform, with country of origin limited to Australia, Canada, New Zealand, UK and USA. Participants self-reported English as a primary language, first language, and fluent language, and had an average acceptance rate $>90\%$ on the platform. Participants received GB\pounds3.38 for an average experiment duration of 24 minutes. Data collection stopped once we obtained full datasets that passed quality thresholds ($>90\%$ accuracy, median response time $<3000$ ms) from 48 participants. A total of 60 participants completed the experiment; 12 were excluded from analysis for failing to pass quality thresholds. A further 50 participants withdrew or failed to complete the study. 
Of the 48 participants included in the analysis, 23 identified as female, 23 male, and 2 non-binary, with reported age groups ranging from $20-29$ to $70-79$. See  \iflabelexists{appendix:demographics}
  {\cref{appendix:demographics}}
  {the Appendices file of the supplemental materials} for detailed demographics. 
The study was approved by the local ethics research board.
\subsection{Procedure and Apparatus}
Participants were redirected from the recruiting platform to our custom-built experimental website. They provided consent and carried out setup instructions: their eyes had to be comfortably and consistently between 40cm and 80cm from the screen, the browser window maximized and at 100\% magnification, the computer plugged in, and no other software or tabs running. The experimental software verified a screen refresh measure of approximately 60fps or excluded the participant. A virtual chinrest calibration procedure\diffold{ ensured}\diffnew{ measured visual subtended angle per pixel, allowing us to render stimuli at }equivalent angular sizes for all participants (see~\cite{li_controlling_2020}). \diffnew{Participants had to retry the calibration procedure if they were not estimated to be within 40-80cm from the screen.}  
Following calibration, participants filled out a short demographic questionnaire and then performed the 500 trials. Participants could take breaks after every five trials.

\subsection{Stimuli}
We compiled 25 stimulus lists, each consisting of 10 words and 10 pseudowords. 
The word selection procedure and lists are in \iflabelexists{appendix:wordlist}
  {\cref{appendix:wordlist}}
  {the supplemental materials}.
The lists were exactly balanced on number of segments (phonemes and syllables) and word length, and approximately balanced on the logarithm of word frequency (using CELEX frequency measures~\cite{baayen_r_h_celex2_1995}). The order in which stimulus lists were presented was fixed across participants. Counterbalancing of condition order therefore also counterbalances for stimulus list. Counterbalancing for stimulus list ensures that all effects of properties of words (e.g., phonological neighbourhood density) result in noise rather than confounds.
The order of stimuli within each list was pseudorandomized for each participant to ensure unpredictability of correct lexical decision answers.

We obtained phoneme and syllable segmentations by consulting phoneme and syllable boundaries provided in online dictionaries. Then we aligned the letters in each word with the phonemes. Some groups of letters like "ough" map to one phoneme while some letters like the silent-e map to no phoneme. Letters that combine to map to a phoneme were grouped together as a phoneme orthographic unit. We obtained segmentations for pseudowords by analogy with similar words. The word list and all associated segmentations are in the preregistration.

We created meaningless phoneme and syllable segmentations for words and pseudowords by manually selecting segment boundaries to disturb the correct segments while retaining the same number of segments as the corresponding meaningful variants.

\subsection{Analysis}

\diffold{
We used Bayesian models with a metric predicted variable (base-10 logarithm of reaction time), one or more nominal factors, depending on the test, and a non-interacting random factor (participant). 
We modeled log-transformed completion time as Student-t distributions, with location determined by the sum of the effects of the factors, which were modeled as normally distributed with gamma-distributed variance. 
We only analyzed correct responses to word stimuli.
}
\diffnew{
We analyzed log-transformed response time using Bayesian models specific to each hypothesis. The model definitions for each hypothesis are provided in \iflabelexists{appendix:hypotheses}
  {\cref{appendix:hypotheses}}
  {the supplemental materials}. Excepting an initial comparison between words and pseudowords (H0.1, H0.2), we only analyzed correct responses to word stimuli. 
The experiment was designed to test time at high accuracy; therefore we do not report error rates. Nevertheless, we checked that error rates were low ($<4\%$) and that there were no statistically detectable differences in error rate between conditions. A report of this analysis is in the supplemental materials.}

\subsection{Results}
The report of the results is organized around the research questions from Section~\ref{sec:exp1:RQs}. When appropriate, we cite specific hypotheses (e.g., H1.1), which refers to the definition and results in the Appendix. 

In addition, we also tested whether responses to word stimuli were faster than responses to pseudowords, both with and without application of interventions. This is a robust empirical result often replicated in the lexical decision task literature~\cite{Rossmeissl1982Identification}. Our analysis offers strong support for both tests ($>99.9\%$ probability). We observed decision time differences between words and pseudowords between 49-66 ms\removed{ ($d=0.26-0.35$)}, in line with previous work~\cite{Rossmeissl1982Identification}. This supports that our study procedures are compatible with previous measures and sufficiently statistically powerful to discern differences of that magnitude.

\inlinesection{RQ1:}\textbf{\textit{Can word identification be enhanced by visually encoding phonologically meaningful segmentation information?}}
We tested ten hypotheses expecting each meaningful segmentation intervention to perform faster than the no intervention condition (H1.1), and all were strongly rejected. Phonologically meaningful visual segmentation impeded rather than enhanced Word identification. \diffold{Figure~\ref{fig:mfulVNone} shows the estimated probable means for each test of intervention, with response time differences relative to no-intervention and associated effect sizes.}\diffnew{Figure~\ref{fig:E1AllPlots} (top panel)} shows the estimated probable means for each test of meaningful intervention and the response time differences relative to no-intervention.

\ifdiffversion
\begin{figure}
\diffoldfig{
    \centering
    \includegraphics[width=\linewidth]{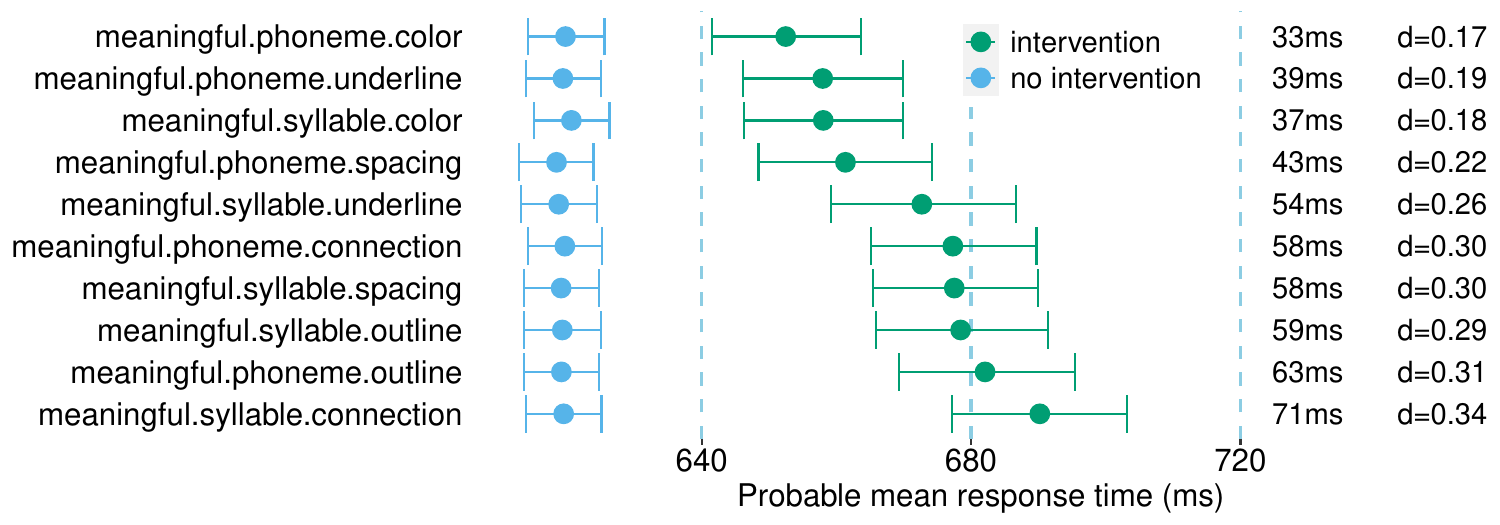}
    \caption{Meaningful Interventions Vs No-Intervention.
    Rows show estimated mean response in tests of each intervention with no intervention. Bars show 95\% HDI, with dots at mean. Printed values are mean difference and effect size.}
    \label{fig:mfulVNone}
}
\end{figure}
\fi

\begin{figure}
\diffnewfig{
    \centering
    \includegraphics[width=\linewidth, alt={Figure shows three panels: No-intervention versus meaningful, no-intervention versus meaningless, and meaningful-meaningless contrast. 95 percent high density intervals  are plotted as horizontal bars for each parameter estimated, with points at the means. Meaningful and meaningless interventions are all clearly slower and not overlapping estimates of no-intervention.}]{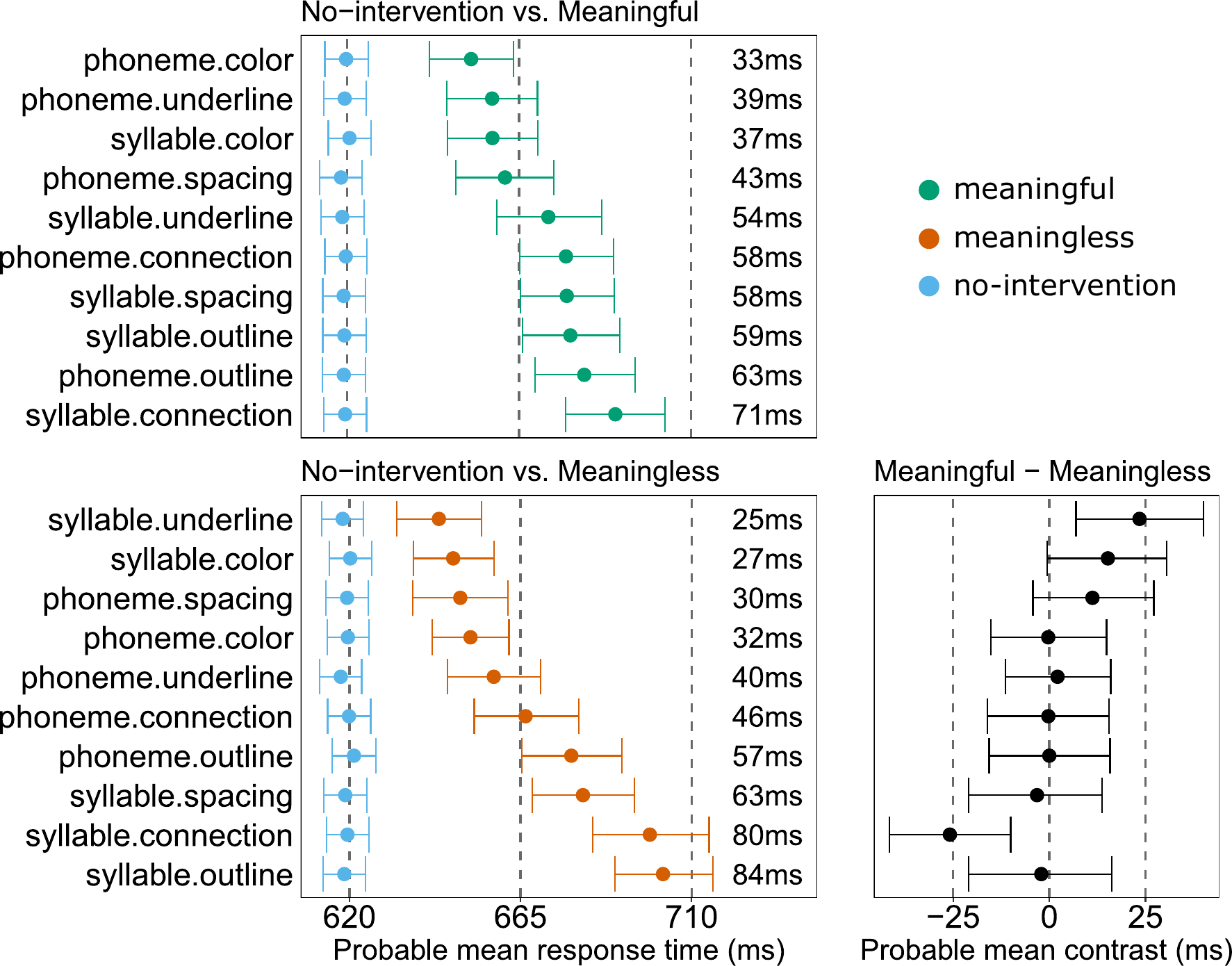}
    \caption{Rows in left panels show estimated time responses per model. The right panel shows estimated meaningful-meaningless contrasts. Bars show 95\% HDI, with dots at mean. Numbers show mean difference.}
    \label{fig:E1AllPlots}
    }
\end{figure}

\inlinesection{RQ2:}\textbf{\textit{Is the information communicated by visual segmentation responsible for any effect on word identification?}}
This question addresses the possibility that effects of interventions are due purely to the change in the appearance of the word, rather than to the (potentially useful) segmentation information made explicit. Therefore, the main hypothesis addressing this question (H2.1) compares performance between meaningful and meaningless segmentations.

Although we hypothesized an advantage of meaningful segmentations, our analysis instead estimates a difference between meaningful and meaningless conditions of between -2.36ms and 8.11ms, which is not conclusive but clearly biased towards meaningful conditions taking longer to process than meaningless conditions (probability of 0.859).

As a follow up, we separate tests of differences between meaningful and meaningless segmentations per the graphical augmentation and the unit of segmentation (syllable or phoneme). Since different graphical techniques and different units might be more or less effective, we expected to see differences between the different combinations of \emph{segmentation type} and \emph{visual design} in terms of how effectively each communicate the information (H2.2).

 H2.2 is strongly supported for syllable.outline, strongly rejected for syllable.underline and weakly rejected for phoneme.spacing. This suggests that the effect of meaningfulness on word identification differs by visual design and grain size.\diffold{ Figure~\ref{fig:mfulVmless}}\diffnew{ Figure~\ref{fig:E1AllPlots} (lower-right panel)} contrasts responses to meaningful and meaningless variants of each intervention.

\ifdiffversion
\begin{figure}
\diffoldfig{
    \centering
    \includegraphics[width=\linewidth]{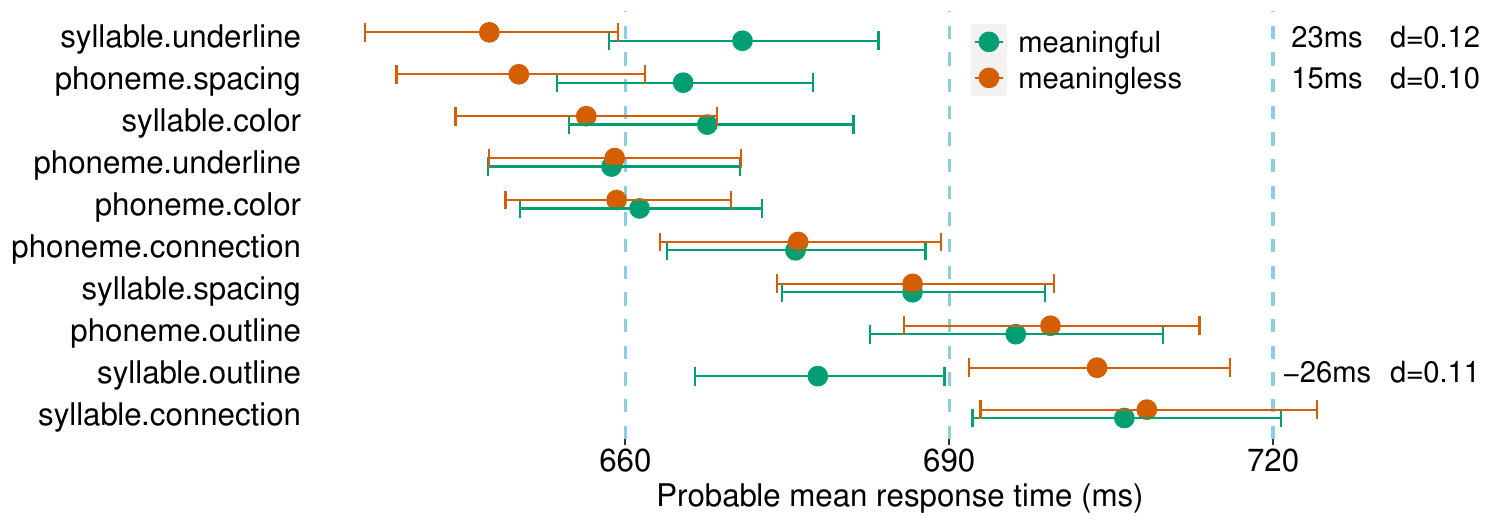}
    \caption{Meaningful Vs.\ Meaningless Variants of Interventions. Rows show estimated mean response in tests of meaningful with meaningless variants of each intervention.}
    \label{fig:mfulVmless}
    }
\end{figure}
\fi

\inlinesection{RQ3:} \textbf{\textit{Does the phonological granularity of visually presented segmentation information affect word recognition performance?}}
As syllables are comparatively more consistent~\cite{ziegler_reading_2005}, they provide more information (reduce more uncertainty) than phonemes. We expected syllables to provide faster performance (H3.1). This hypothesis is strongly rejected ($p<0.001$), indicating that phonological granularity does affect word identification, but not in the expected direction: phoneme segmentation outperforms syllable segmentation, with a mean difference of 8.81ms\removed{ (d=0.05)}. 

\inlinesection{RQ4:} \textbf{\textit{ Can word identification be impeded by phonologically meaningless visual segmentation?}}
We hypothesized that meaningless interventions would slow down the task compared to no intervention or, in other words, that any graphical manipulation of text that does not provide information would have a cost (H4.1). This hypothesis is strongly supported for all meaningless interventions (all $p>.999$) and average time costs range between 25.33 and 83.89ms with syllable.underline the least costly and syllable.outline the most interfering graphical intervention. \diffold{Figure~\ref{fig:mfulVNone}} \diffnew{Figure~\ref{fig:E1AllPlots} (lower-left panel)} shows the estimated probable means for each intervention.

\ifdiffversion
\begin{figure}
\diffoldfig{
    \centering
    \includegraphics[width=1\linewidth]{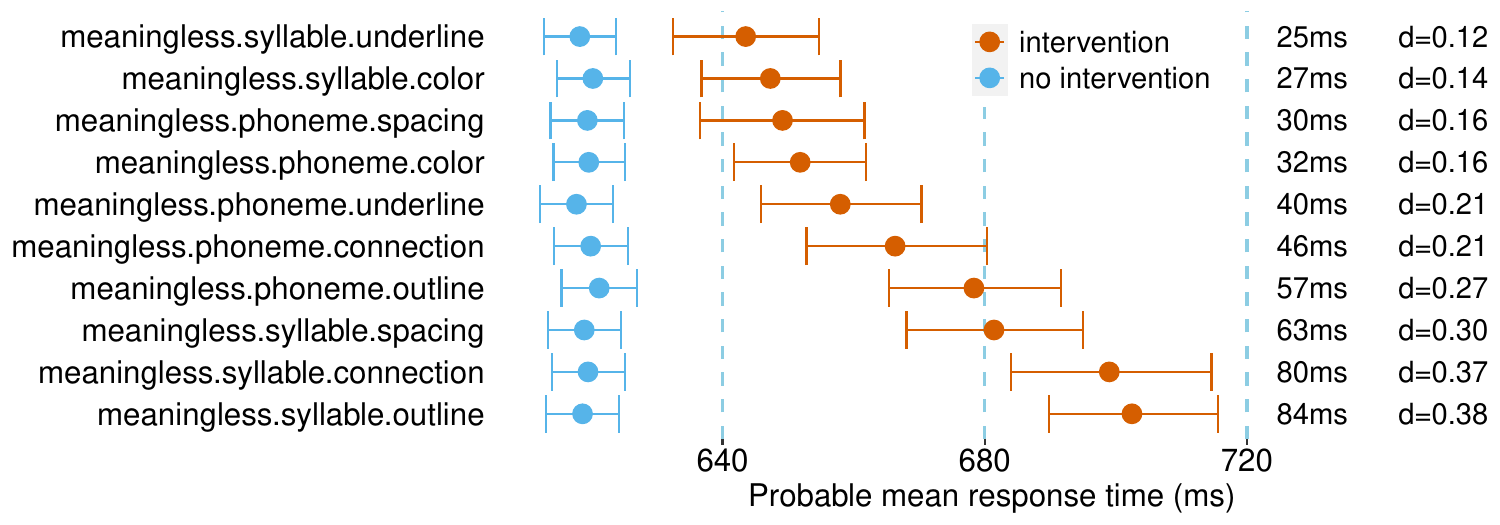}
    \caption{Meaningless Interventions Vs.\ No-Intervention. Rows show estimated mean response in tests of each intervention with no intervention.}
    \label{fig:mlessVnone}
    }
\end{figure}
\fi

\inlinesection{RQ5:} \textbf{\textit{ Which interventions have a greater or lesser impact on word identification?}}
We tested pairwise comparisons between all meaningful interventions (H5.1). We found strong differences between many interventions, in the range of 14.40-40.16ms.
Similarly, we tested pairwise comparisons between meaningless interventions (H6.1) and found strong differences between many interventions, in the range of 18-56ms.
The conclusive contrasts suggest a similar grouping for meaningful and meaningless conditions. The most favorable interventions were phoneme.color, phoneme.underline, syllable.color, phoneme.spacing, and syllable.underline. The slowest were syllable.connection, phoneme.outline, syllable.spacing and syllable.outline.
\diffnew{
Besides the pre-registered analyses, we conducted an exploratory analysis to test for participant fatigue, of which we found no evidence (a \emph{Fatigue} report appears in the supplemental materials).}

\subsection{Experiment 1 Discussion}

All interventions, meaningful or meaningless, impede word identification by between 25ms\removed{(d=0.12)} and 84ms\removed{(d=0.38)}. This represents a large disruption, given that empirical consensus estimates the processing time from stimulus presentation to lexical access in the range of 175-200ms~\cite{rayner2012psychology}, and that the difference between high and low frequency words, considered to be a large effect in psycholinguistics literature, is estimated at around 30ms~\cite{rayner2012psychology}. Furthermore, the differences observed between interventions suggest that choice of visual encoding has an appreciable impact on word identification performance, with differences of up to 56ms.

A surprising result is that segmentation information improves performance relative to meaningless controls with syllable.outline, but impedes performance with phoneme.spacing and syllable.underline (see \diffold{Figure~\ref{fig:mfulVmless}} \diffnew{Figure~\ref{fig:E1AllPlots}, lower-right panel}). Why should the same information improve or impede performance depending on the graphical qualities of the intervention? The graphical qualities of syllable.outline are more disruptive (as measured by performance of the meaningless controls) than phoneme.spacing or syllable.underline (by 52-56ms). Results for syllable.outline may then be analogous to the phenomenon reported by Katz and Baldasare~\cite{katz_syllable_1983} in which meaningful segmentation only outperforms meaningless controls in adverse settings, such as when word identification is made difficult by alternating letter case. However, the finding that segmentation information impedes performance for the less graphically disruptive interventions is new and unexplained.

Previous studies (see Section~\ref{sec:relatedSegmentationStudies}) have interpreted a difference between meaningless and meaningful variants as an effect of disturbing segment boundaries. However, this interpretation does not explain faster responses for meaningless variants. It also fails to explain why we observe a difference for syllable.outline but not syllable.connection, when both meaningful variants preserve segment boundaries.

A better interpretation is that meaningful interventions communicate segment boundaries, and that word identification with this information is slower in normal circumstances, but less sensitive to graphical disturbance. We would then expect that, provided they successfully communicate the segmentation information, meaningful interventions should impede performance if their meaningless controls are not very disruptive, enhance performance if their controls are very disruptive, and show no effect if their controls have more intermediate performance. This is compatible with our results and prior related work. Additionally, the results for syllable.connection can be explained as a failure to effectively communicate the information.

 The within-subject study design could have influenced the results: participants had to carry out tasks where segmentation information could be either meaningful or not, often in subsequent blocks, perhaps encouraging them to ignore the signal, conciously or unconciously. This is a reasonable strategy when signals are noisy or inconsistent. 

 Participants may not have had sufficient training with the unfamiliar typographic interventions to benefit from them. Explicit instruction regarding the meaning and use of the interventions may also be necessary, as has been previously found with related typographic interventions (see Section~\ref{sec:relatedWork:enhancingTypographicVariables}). Without instruction, participants may also have organically adopted different strategies with respect to the interventions: to ignore or use the information they present, which may yield heterogeneous results.
\section{Experiment 2: Effects of Training with Visual Segmentation on Lexical Decision}
To address the points mentioned in the Experiment 1 Discussion regarding the influence of the study design and lack of training, we designed another experiment.\footnote{The pre-registration of this study is available at \url{https://doi.org/10.17605/OSF.IO/GC7QX}} We used the same forced-choice lexical decision paradigm, but with a between-subjects design and an extended training block with the assigned intervention. 

With full datasets from 48 participants, we did not have sufficient statistical power to detect effects of the size expected from Experiment 1. We also needed to make some significant changes from the preregistered analysis plan, as the original models were unable to adequately account for large between-group baseline imbalances.
We therefore refrain from reporting the results here, although the results informed our design of Experiment 3. A small summary of the results and power analysis is available in the supplemental materials.
\section{Experiment 3: Effects of Training Strategy on Lexical Decision with Visual Segmentation}
\diffold{Informed by Experiment 2, and to additionally address the points in the Experiment 1 Discussion regarding explicit instruction, we designed another experiment using a forced-choice lexical decision paradigm, and included an extended training block with the assigned intervention. We added a follow-up training task after each lexical decision task in the training block, which was designed to encourage either using or ignoring the visual segmentation information in the lexical decision task. We reasoned that the difference in performance between the two strategies would indicate an effect of the segmentation information, provided the training was successful. Immediately before the training block, we also included an instructional video explaining the meaning of the visual encodings, and explicitly instructing the participant to adopt their assigned training strategy. We selected two of the interventions from Experiment 1: syllable outline and syllable underline, as these showed strong but opposing effects of segmentation information.}

\diffnew{The experimental design of Experiments 1 and 2 left open some possible reasons why the evidence contradicts our hypotheses. We designed Experiment 3 to address four main possibilities: A) The within subject design, where each participant was exposed to both meaningful and meaningless segmentations, could have removed or diluted the participant's ability or incentive to use the segmentation information because, half of the time, using the information would be counterproductive; B) Participants were not trained to use the interventions and the interventions may help only when explicitly learned; C) Participants do not have an incentive to pay attention to the segmentation, and might ignore it all together, and; D) Participants might not have been exposed to sufficient training to use the interventions effectively.}

\diffnew{The design of Experiment 3\footnote{The pre-registration of this study is available at \url{https://doi.org/10.17605/OSF.IO/F5397}.} is still based on the lexical decision task, but addresses the points above by using a between-subjects design in which each participant would only be exposed to meaningful interventions (A), by providing participants explicit instruction on how the interventions work (B), by introducing a secondary task after each trial that requires paying attention to the segmentation (C), and by reducing the number of interventions to two, hence multiplying the exposure of each participant to the same intervention (D).}

\subsection{Tasks \diffold{and Measurement}}\label{sec:exp3:tasks}
\diffold{
The lexical decision task was identical to Experiment 1 (see~\ref{sec:exp1:task}). In the training block, participants completed a training task after each lexical decision task. In the training task, the participant was shown text that was part of the preceding lexical decision stimulus (target) or not (distractor). The participant responded by pressing the ``A'' key if they thought it was part of the stimulus, or ``L'' if not. We measured elapsed time from stimulus onset and response accuracy. The training task stimulus was rendered in an identical font and size to the lexical decision stimulus, but with a slight vertical offset to make the two tasks visually distinct. An instructional prompt appeared below the stimulus.}

\diffnew{
The lexical decision task was identical to Experiment 1.
In the training block (see Section~\ref{sec:exp3:procedure}) participants completed a secondary task after each lexical decision task that we call the \emph{training task}. In this extra task the participant saw a sequence of letters that might have been part of the word shown in the lexical decision task. The participant chose whether the sequence was part of the word by pressing the ``A'' key, or not, by  pressing ``L''. For example, if the lexical decision stimulus was ``possible'', the training task might show ``ble'', for which the right answer was pressing the A key (target). In half the cases, a letter was changed so that the sequence did not match the previous stimulus (e.g., ``bje''), in which the correct answer was ``L''. 
}

\diffnew{
The new task introduces a new between-subjects factor (\emph{Training strategy}), with two conditions: \textbf{attend} and \textbf{ignore} (Figure~\ref{fig:taskTiles}). In the \emph{attend} condition, the sequence in the training task always corresponded to a syllable as segmented in the LD task stimulus (``ble'', or ``pos'' or ``si'' in the example from Figure~\ref{fig:taskTiles}), whereas in the \emph{ignore} condition the sequence crossed syllables. This design makes it so that taking into account the segmentation from the LD task stimulus facilitates identification of the sequence in the training task for the \emph{attend} condition, but impedes identification for \emph{ignore} because the matching would have to contend with crossing between syllables.}

\begin{figure}[h]
  \diffnewfig{
    \centering
    \includegraphics[width=1.0\linewidth, alt={Figure shows the sequence of stimuli for a lexical decision task followed by a training task. Examples are shown for target and distractor segment types, for both the attend and ignore training conditions.}]{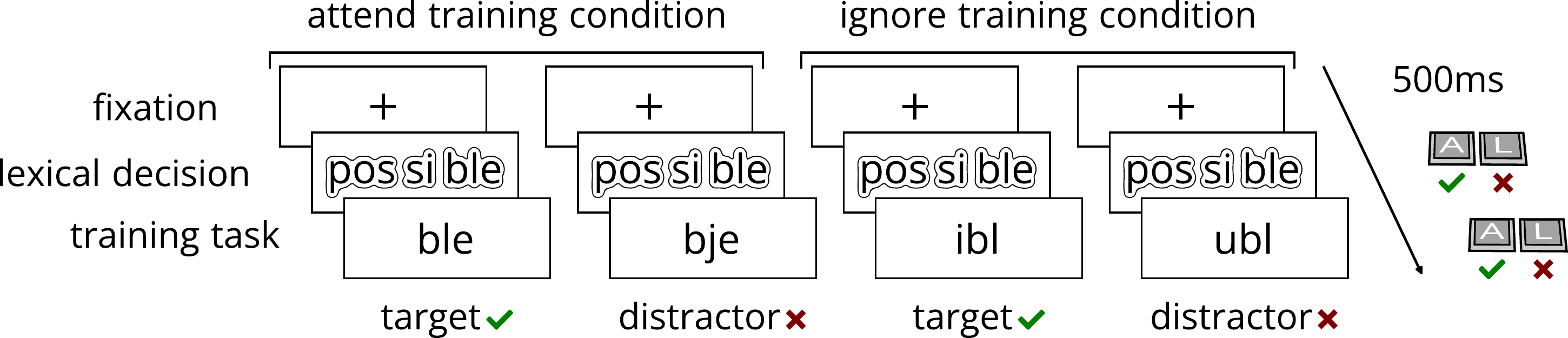}
    \caption{Target stimuli (tiles) for the LD and training task in Exp. 3.} 
    \label{fig:taskTiles}
  }
\end{figure}

\subsection{Experimental Factors}
The experiment manipulated three factors and logged which block each trial belonged to. \emph{Stimulus type} (word or pseudoword) works as in Experiment 1. The other variables are:

\diffold{\inlinesection{Visual Segmentation Design}
The way in which the different segments are visually differentiated from each other (\textbf{underline, outline}). See Section \ref{sec:visualDesign} for a description of each.}
     
\diffnew{
\inlinesection{Visual Segmentation Design (between subjects)}
We selected the \textbf{underline} and \textbf{outline} at the grain of syllable because they exhibited the most extreme opposing effects in Experiment 1.}

\diffold{
\inlinesection{Training Strategy}
Whether the training and instructions were designed to encourage attending to or ignoring the segmentation information (\textbf{attend, ignore}). In the \emph{attend} condition, training task target stimuli corresponded to one of the segments in the lexical decision stimulus, whereas target stimuli crossed segment boundaries in the \emph{ignore} condition. This makes the segmentation information provided in the lexical decision stimulus beneficial or counterproductive to the training task, providing an incentive to attend to or ignore the segmentation intervention.}
    
\diffnew{    
\inlinesection{Training Strategy (between subjects)}
    Whether the secondary task encouraged or discouraged attention to the segmentation stimuli from the LD task (\textbf{attend} or \textbf{ignore}---see Section~\ref{sec:exp3:tasks}).}     
    
\inlinesection{Block Type}
    Whether the trial occurred in the test block before or after the training block (\textbf{pretest} or \textbf{post-test}).

\inlinesection{Segment Type (training task only)}
    Whether the training task stimulus is part of the associated lexical decision stimulus (\textbf{target} or \textbf{distractor}). This is needed to make the training task non-trivial.

\subsection{Experimental Design and Conditions}\label{sec:exp3:procedure}
\diffold{We used a mixed factorial design. Participants completed a total of 300 lexical decision trials and 120 training task repetitions (see Figure \ref{fig:E3Blocking}). Only data from the 160 pre-test and post-test lexical decision repetitions (80 with the assigned intervention, 80 with no intervention) were used in the analysis.} 

\diffold{Participants were assigned to one of the four combinations of training strategy with visual design (the \emph{intervention}), excluding the \emph{none} or \emph{no intervention} condition. }

\diffnew{Participants were assigned to one of four combinations of training strategy and visual segmentation design. Participants completed a total of 300 lexical decision trials and 120 training task trials in three blocks (see Figure~\ref{fig:E3Blocking}): a pre-test (without training task, 40 trials with intervention, 40 trials without), a training block (120 pairs of LD + training tasks), and a post-test (without training task, 40 tasks with intervention, 40 without). The main analyses consider only trials in the pre-test and post-test phases; the training block effects manifest through differences in the post-test.}

\begin{figure}[h]
  \diffnewfig{
    \centering
    \includegraphics[width=0.95\linewidth, alt={Figure shows the sequence of block types with number of repetitions in Experiment 3. The sequence is: 20 lexical decision (LD) familiarization with no-intervention, 40 LD pre-test each counterbalanced with and without intervention, 120 LD interleaved with training task with intervention, and 40 LD post-test each counterbalanced with and without intervention.}]{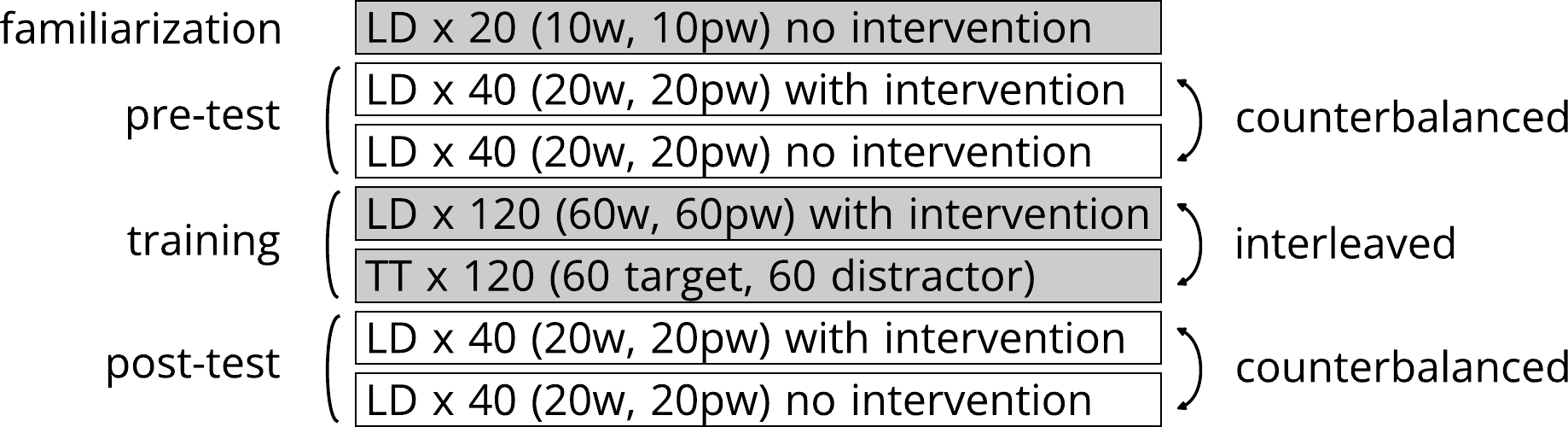}
    \caption{Blocking of repetitions of lexical decision (LD) and training (TT) tasks. Note w=word and pw=pseudoword.}
    \label{fig:E3Blocking}
    }
\end{figure}

We counterbalanced assignment of stimulus lists to each of the four conditions in the test blocks: 2 (intervention vs.\ none) $\times$ 2 (pre-test vs.\ post-test) using a digram-balanced Latin square. We also counterbalanced for condition order by including each of the four possible condition orders in the test blocks: 2 (intervention first vs.\ none first) $\times$ 2 (pre-test vs.\ post-test), for a total of 4 $\times$ 4 $= 16$ counterbalance groups. Six participants were assigned to each counterbalance group. \diffnew{One participant from an attend underline group was excluded from analysis (see Section \ref{sec:E3participants}).}
 
\subsection{Research Questions}

We focused on the following research questions (RQs):\footnote{RQ references in the paper and the pre-registration differ for clarity.}
\begin{description}
\item[\texttt{RQ1}:] Can meaningful segmentation enhance word identification with explicit instruction and training to attend to the presented information?
\item[\texttt{RQ2}:] If segmentation information can be both ignored and attended to, does the information enhance or impede word identification?
\item[\texttt{RQ3}:] Can meaningful segmentation enhance word identification without training?
\item[\texttt{RQ4}:] Which \emph{visual design} most enhances or least impedes word identification?
\item[\texttt{RQ5}:] With training, do participants adapt to the presence of the visual segmentation interventions?
\end{description}

\diffold{As with Experiment 1, we do not report results for each tested hypothesis in the text and provide the hypotheses and corresponding test results in}
\diffnew{Comprehensive results for all tested hypothesis are in} \iflabelexists{appendix:hypotheses}
  {\cref{appendix:hypotheses}}
  {the Appendix file in the supplemental materials}.

\subsection{Participants}\label{sec:E3participants}
We recruited participants ($>18 yo$) through Prolific with $>95\%$ on the platform from Australia, Canada, New Zealand, UK and USA. We excluded participants from the previous experiments. Participants reported English as a first language, primary language, fluent language, and earliest childhood language. Participants received GB\pounds3.31, with an average completion time of 23.5 minutes.

Data collection stopped once we obtained full datasets passing quality thresholds ($>85$\% accuracy, median response $< 3000$ms in lexical decision trials; $>75$\% accuracy, median response $<3500$ms in follow-up training tasks) from 96 participants. 132 participants completed the experiment, of which 36 were below the thresholds. A further 30 participants failed to complete the study. \diffnew{One of the 96 participants was found to have partially attempted the experiment before retrying and completing it (we removed all data from this participant)}. The study was approved by the local research ethics board.

Of the 95 participants analyzed, 36 identified as female, 57 male, and 2 non-binary, with age groups ranging from $<20$ to $70-79$. Detailed demographics are in \iflabelexists{appendix:demographics}
  {\cref{appendix:demographics}}
  {the Appendix in the supplemental materials.}

\subsection{Procedure and Apparatus}
All steps leading up to the lexical decision tasks were identical to Experiment 1, except that participants with screen refresh rates greater than 60fps were not discarded. The lexical decision task was identical to Experiment 1.

After the participant completed the first 100 lexical decision trials (the task training and pre-test blocks), they watched an instructional video specific to their assigned condition which explained that the additional markings grouped letters into syllables, described the follow-up task procedure, specified whether the follow-up letter sequences correspond to syllables or cross syllable boundaries, and instructed the participant to pay attention to or ignore the markings. Participants then performed the 240 trials (120 lexical decision, 120 training task) in the training block, followed by the remaining 80 lexical decision trials.

\subsection{Stimuli}
We compiled 4 balanced stimulus lists for the test blocks (pretest and post-test), each consisting of 20 words and 20 pseudowords. The lists were exactly balanced on number of segments (phonemes and syllables) and word length and approximately balanced on the logarithm of word frequency (using CELEX frequency measures). The words and pseudowords in the training were exactly balanced on length and number of segments. As in Experiment 1, stimuli order within each list was pseudorandomized per participant. We re-used the meaningful and meaningless syllable boundaries defined in Experiment 1.

We prepared target and distractor strings for attend and ignore conditions for each word and pseudoword in the training stimulus lists. Distractor strings were created by replacing up to half of the letters in the corresponding target string, while ensuring that the result is not part of the corresponding lexical decision stimulus, but is part of at least one word in the word list (and is therefor not easily identified as an illegal letter sequence in English). 
For each participant, half of the words and half of the pseudowords in the training block were pseudorandomly assigned to have their corresponding target presented in the training task, while the remainder were assigned to have the distractor presented, ensuring unpredictability of correct answers.

\subsection{Analysis}
\diffold{We used Bayesian models with a metric predicted variable and one or more nominal factors, depending on the test, as well as a non-interacting random factor (participant). Depending on the test, the predicted variable was the base-10 logarithm of response time ($log\_rt$), or $log\_rtz=log_{10}(rtz+offset)$ where per-participant normalized reaction time ($rtz$) is defined as the difference between the response time and the per-participant median response time in the no intervention condition in the corresponding block type (pretest or post-test), and the per-test offset ($offset$) is calculated for each test such that $log_{10}(rtz+offset)>=0$ for all data relevant to the test. We found in Experiment 2 that this normalization was necessary to accommodate large between-group differences in participant variance.}

\diffnew{We used Bayesian models specific to each hypothesis. The model definitions for each hypothesis are provided in \iflabelexists{appendix:hypotheses}
  {\cref{appendix:hypotheses}}
{the Appendix included in the supplemental materials}. Tests involving a single intervention and the no intervention condition used log-transformed response time as dependent variable, and were treated as purely within-subjects by filtering the data to only the participants assigned to the particular intervention. For between-subjects tests (such as between interventions) the dependent variable was the log-transformed per-participant response time change score, defined as the difference between the response time and the per-participant no-intervention median response time in the corresponding block type (pre-test or post-test). This is often referred to as a \emph{change score} analysis~\cite{allison1990change_scores}. We assume that the distribution shape of time change scores is similar to other time measures; however, it might be occasionally negative, and hence we add a constant before transforming.\footnote{We carried out a sensitivity analysis that confirms that a range of values of this constant do not affect the comparisons in a significant way. See the \emph{Sensitivity} report in the supplemental materials.}
A more detailed description of the analysis is in \iflabelexists{appendix:hypotheses}
  {\cref{appendix:hypotheses}}
  {the Appendix included in the supplemental materials}.
}

\diffold{We modeled the independent variable ($log\_rt$ or $log\_rtz$) as Student-t distributed, with location determined by the sum of effects of the nominal factors, which were modeled as normally distributed with gamma-distributed priors for the variance. We only analyzed correct responses to word stimuli.}

\diffold{Tests involving a single intervention and the no intervention condition used $log\_rt$ as dependent variable, and were treated as purely within-subjects by filtering the data to only the participants assigned to the particular intervention. For between-subjects tests (such as between interventions), we used $log\_rtz$ as dependent variable, and we separated participant factors between conditions.}

Where applicable, we estimate and report the minimum parameter value that we would have detected with the present amount of statistical power. With 95\% probability, the true parameter value is within the 95\% HDI of the posterior. If the true value were at least the length of the 95\% HDI away from zero, then with 95\% probability, the HDI would exclude zero, providing a conclusive result. 
\diffnew{As in Experiment 1, errors rates were low ($<3\%$) and indistinguishable between conditions.}

\subsection{Results}

As with Experiment 1, we first tested whether responses to words are faster than responses to pseudowords, without and with interventions, before and after training respectively (H0.1-0.4). All hypotheses were strongly supported with an estimated probability $>0.999$. Word-pseudoword effects were larger than observed in Experiment 1, with mean differences of 110.05ms, 153.95ms, 95.61ms and 93.12ms respectively, but are within ranges found in previous work~\cite{Rossmeissl1982Identification,rayner2012psychology}. Results support compatibility of our study procedures with previous measures and sufficient statistical power to discern differences of that magnitude.


\inlinesection{RQ1:} \textbf{\textit{ Can meaningful segmentation enhance word identification with explicit instruction and training to attend to the presented information?}} We tested two separate hypotheses (H1.1, H1.2) that each \emph{visual design} would perform faster than no intervention after training to attend to the segmentation information. This hypothesis was strongly rejected for both underline (22.69ms difference) and outline (54.41ms difference), indicating that both interventions impede word identification, even with explicit instruction and training. 

\inlinesection{RQ2:} \textbf{\textit{ If segmentation information can be both ignored and attended to, does the information enhance or impede word identification?}}
We reasoned that, if the segmentation information is communicated when attended to but not when ignored, and if the training strategies successfully encourage attending and ignoring respectively, then the order of post-training response times between training strategies, relative to each-other and the no intervention condition, would provide evidence for a facilitatory or inhibitory effect of the information on word identification.

We found strong indication that responses with visual segmentation interventions were slower than with no intervention, for either \emph{visual design}, when trained to ignore (H2.2, H2.5) or attend (H2.1, H2.4). Slowdowns (relative to no intervention) when trained to ignore, for underline and outline respectively, were estimated at 16.57ms and 47.66ms. 
We found no conclusive difference between the ignore and attend strategies (H2.3, H2.6), suggesting that either the training strategies failed to differentiate behaviour, or the effect of the communicated information was too small for tests to detect. Post-hoc power analyses indicate that we would have detected a difference of at least 24.75ms for underline, or 29.26ms for outline, with a probability of 95\%. 

\inlinesection{RQ3:} \textbf{\textit{ Can meaningful segmentation enhance word identification without training?}} The analysis strongly supports that meaningful segmentation impedes lexical decision before training, for either \emph{visual design} (H3.1, H3.2), corroborating Experiment 1 findings. Underline and outline designs increased response times relative to no-intervention by an estimated 19.34ms and 63.44ms respectively.

\inlinesection{RQ4:} \textbf{\textit{ Which \emph{visual design} most enhances or least impedes word identification?}}
We tested three hypotheses on the difference between response times with the underline and outline \emph{visual designs}: when trained to attend (H4.1), when trained to ignore (H4.2), and without training (H4.3). All results were in the direction of faster responses for underline, with the hypotheses strongly supported without training (20.7ms difference), weakly supported when trained to ignore (11.9ms difference), and not conclusive when trained to attend. 


\inlinesection{RQ5:} \textbf{\textit{ With training, do participants adapt to the presence of the visual segmentation interventions?}} We hypothesized that participants would improve at the lexical decision task (performance with no intervention) between the pretest and post-test blocks (H5.1); this was strongly supported, with an estimated mean difference of 6.29ms. We further separately hypothesized that participants would improve in the presence of each intervention, when trained to attend (H5.2, H5.3) or ignore (H5.4, H5.5); all tests were in the hypothesized direction, but were only conclusive for ignore.outline (strongly supported, with a mean difference of 20.89ms) and attend.underline (weakly supported, mean difference of 15.89ms). We also hypothesized that post-training performance gains would be larger with each intervention and each training strategy than with no intervention (H5.6--H5.9), which we interpret as improvement beyond familiarization with the task; this was only supported (strongly) for ignore.outline (20.66ms difference). 


\subsection{Experiment 3 Discussion}
We found that, even with instruction and training, phonological segmentation impedes word identification. 

We found no conclusive difference between attend and ignore conditions, whereas we would have detected differences of 25ms for underline, or 29ms for outline. These values are very close to the corresponding probable meaningful vs.\ meaningless differences indicated by Experiment 1: 23ms and 26ms. While the contrasts are in the direction predicted by Experiment 1, we would have expected a stronger indication given the calculated power. We could interpret this as evidence against an effect of segmentation information, or that training was not effective in encouraging either attending or ignoring.


Further supporting an interpretation that training was not effective, we only found a training effect beyond that attributed to familiarization with the task itself for ignore.outline.
\section{Discussion}
Our results have implications for researchers and designers of reading enhancement interventions and text visualizations.

\subsection{Implications for Reading Enhancement}
Our primary goal was to assess whether sublexical graphical segmentation interventions can enhance word recognition for normal skilled readers. We have found that it is instead disruptive, even after a certain amount (120 trials) of dedicated training and explicit instruction (at least for our five graphical designs segmenting phonemes or syllables). 

We expected meaningful variants of interventions to outperform meaningless ones, given the results of prior related work (see Section~\ref{sec:relatedSegmentationStudies}). The finding that segmentation information can enhance or impede lexical decision (relative to meaningless controls) depending on the graphical intervention is novel and surprising. We propose an interpretation that accounts for our data and the findings of prior studies: word identification with segmentation information is slower for normal text, but is less sensitive to graphical disturbance. In contexts of high graphical disturbance, such as with separator marks (e.g., //)~\cite{treiman1994extent}, our outline designs, and when text is printed in alternating case~\cite{katz_syllable_1983}, the information is beneficial. At more intermediate levels of graphical disturbance, such as segmentation designs using case~\cite{katz_linguistic_1981,bowey1996orthographic} or our spacing design, we observe little difference between meaningful and meaningless variants.

Our interpretation makes testable predictions that future work should investigate. We expect performance of meaningful interventions to improve, relative to meaningless controls, when graphical disturbance is increased (e.g., by presenting text in alternating case or applying spatial noise). Some interventions may outperform no-intervention controls in such conditions, which might suggest promise as a reading enhancement for impaired vision.

Training and instruction seems to have little effect, although there is some indication that impacts of highly disruptive designs lessen with training. An experiment that includes both meaningfulness and training strategy is needed to assess whether training strategy is successful.


Readers of today's traditional orthography are disrupted when inter-word spaces are removed~\cite{mirault_reading_2019}, but the introduction of inter-word spaces may have initially impeded readers who were already skilled with \emph{scriptura continua}. While there are mixed results with some unspaced orthographies such as Japanese~\cite{sainio_role_2007} and Thai~\cite{winskel_eye_2009}, there is evidence that introducing inter-word spaces into the traditionally unspaced Chinese writing system is disruptive to skilled readers~\cite{ma_visual_2019}, but helpful for children~\cite{wang_space_2015} and L2 learners~\cite{shen2012eye}. The same is plausible for orthographic segmentation in English. Prior studies involving children show an advantage of meaningful segmentation over meaningless controls~\cite{katz_linguistic_1981}, and VIE evaluations have shown benefits for L2 learners~\cite{lee2006synthesis,lee_visual_2008,han_textual_2008}.

Further research should explore the optimal balance between necessary and excessive visual segmentation in reading materials. This involves not only reconsidering how much visual distinction is beneficial but also identifying which types of text or reader characteristics (such as age, reading skills, or language proficiency) influence the effectiveness of these visual strategies. 

\subsection{Implications for Visualization Design}
Our secondary goal was to compare the impact of different graphical interventions on word identification. We observed large differences in performance between the tested graphical interventions. These results have implications for visualization designers and researchers. Text visualization designers should choose their visual encodings carefully, as their choice will impact word identification and reading. Our results also suggest that visualization techniques that involve adding marks near the text (e.g., sparklines~\cite{jelen_using_2011}) or modifying the text itself (e.g.,\cite{brath2020visualizing,nacenta_fatfonts_2012}) comes at the cost of slowing down recognition. Designers and practitioners should consider whether the value of these visualizations justifies the added reading difficulty. On the positive side, the results can also be used to select the types of interventions that are less intrusive.

In text visualization, \emph{visual marks} (graphical units---what Brath~\cite{brath2020visualizing} terms "scope of text") may be composed of multiple physical marks (glyphs). For example, words can be used as visual marks to represent associated data (e.g., their frequency within a passage). By relating our work to the text visualization literature, we realized that our interventions visually demarcate the groups of glyphs that comprise visual marks for phonological data (e.g., the frequency of each phoneme within a passage). Future work should investigate the articulation of visual marks in text visualization, as it is underexplored in the literature. 

\subsection{Limitations}

Using lexical decision as a proxy for word identification, although commonplace, might not always be predictive of reading performance. It may be possible to answer correctly without identifying the word, for example, by identifying predictive features of words vs.\ pseudowords. Our graphical interventions may also interact with such features. For this reason we have confirmed that our experiments reproduce the robust empirical result often replicated in the lexical decision task literature that responses are slower for pseudowords than words.

While word identification is observed to be similar in isolation and in connected text~\cite{rayner2012psychology}, these established findings are with unaltered words. The tested visual segmentation interventions may interact with reading processes in connected text in ways not predicted by studies of isolated word identification. 
\diffnew{Relatedly, we only measure word identification speed. Future work could use other paradigms such as passage reading to investigate the impacts of visual segmentation on other aspects of reading performance such as comprehension.}

\diffold{We don't account for incorrect responses, and only analyze response latency. We could therefore be missing tradeoffs between speed and accuracy. We implicitly assume near ceiling accuracy. We only analyzed data passing accuracy thresholds, but this also may bias the participant sample, affecting generalizability.}

As we have not taken measures to assess reading ability, our participant sample may be biased in this regard. There is likely some self-selection toward better-skilled readers, as the reward is fixed and the task is more difficult for less skilled readers.

\diffnew{
Finally, although we cannot rule out that substantially more training (e.g., over weeks) could result in performance benefits from visible segmentation, we do not have evidence suggesting this. In fact, our analysis of learning effects shows no statistical difference between the first and second parts of the post-training blocks, suggesting that training effect was saturated (the exploratory analysis is available in the \emph{Training} report of the supplemental materials). }

\section{Conclusion}
We introduced visual segmentation as a technique that uses typographic parameters to make orthographic units (e.g. syllables) within words more conspicuous, which we hypothesized would enhance word identification. We conducted three internet-based experiments testing the effects of visual segmentation interventions on word identification, using a lexical decision paradigm. Our goals were to determine whether visual segmentation interventions can enhance word identification, and which typographic parameters have a greater or lesser impact.

In the first experiment, we tested ten visual segmentation interventions --- five typographic parameters with syllable or phoneme units, as well as a no intervention control condition --- in a fully-crossed within-subject factorial design. The second experiment suffered from low power, but informed design of the third. In the third experiment, we selected two of the five syllable segmentation designs and tested whether effects of the intervention are different with training and instruction.

We found that all tested interventions impede word identification, slowing responses by at least 17ms\removed{ ($d=0.11$)}, even with training and instruction. We also found that choice of visual encoding has an appreciable impact on word identification performance, with differences of up to 56ms\removed{ ($d=0.28$)}. These effects are comparable to response differences between words and pseudowords, which is considered to be a large effect in studies of word identification. Surprisingly, we found that, depending on the \emph{visual design} of the intervention, information communicated by the interventions can either inhibit, enhance, or have no effect (relative to meaninglessly segmented controls). We observed no conclusive difference in responses when participants were trained to ignore or attend to segment boundaries, which we interpret as suggesting that the marks are difficult to ignore, even with training. 

Our findings reconcile ambiguous results of prior related studies, and suggest areas for future research. They are also relevant to designers who want to overload text renderings with additional information, such as in visualization, or for graphical or aesthetic purposes. 


\section*{Supplemental Materials}
\label{sec:supplemental_materials}

All supplemental materials are available on OSF at \url{https://doi.org/10.17605/OSF.IO/79Y5S}, released under a CC BY 4.0 license.
In particular, they include (1) {data, word lists, analysis notebooks, and reporting notebooks used to generate figures for each experiment}, (2) {four additional reports of exploratory analysis} and (3) an appendix to this paper.





\acknowledgments{%
  The authors wish to thank the reviewers for their thoughtful and detailed reviews, which have led us to improve this paper substantially. This project was funded by Canada's NSERC (award 2020-04401).%
}

\bibliographystyle{abbrv-doi-hyperref}

\bibliography{template}

\appendix 

\end{document}